\documentclass[12pt]{aastex}

\usepackage{emulateapj5}
\usepackage{graphicx}

\submitted{Accepted for publication by The Astrophysical Journal}
\shorttitle{\ion{H}{1} shell in the outer Galaxy}
\shortauthors{UYANIKER \& KOTHES}

\begin{document}
\title{A large  atomic hydrogen shell in the outer Galaxy:
SNR or stellar wind bubble?  }
\author{
        B\"ulent  UYANIKER\altaffilmark{1} and
        Roland  KOTHES\altaffilmark{1,2}
}
\altaffiltext{1}{National Research Council,
       Herzberg Institute of Astrophysics,
       Dominion Radio Astrophysical Observatory,
       P.O. Box 248, Penticton, B.C.,
       V2A 6K3 Canada
    }
\altaffiltext{2}{
         Department of Physics and Astronomy,
         The University of Calgary,
         2500 University Dr. NW,
         Calgary, AB,
         T2N 1N4 Canada
 }

\email{bulent.uyaniker@nrc.ca, roland.kothes@nrc.ca}

\begin{abstract}
We report the detection of a ring like \ion{H}{1} structure toward
$\ell$=90$\fdg$0, $b$=2$\fdg$8 with a velocity of $v_{LSR}=-$99 km
s$^{-1}$.  This velocity implies a distance of d=13 kpc, corresponding
to a Galactocentric radius of R$_{\rm gal}=15$ kpc.  The
$\ell-v_{LSR}$ diagram implies an expansion velocity of
$v_{exp}\simeq15$ km s$^{-1}$ for the shell.  The structure has an
oblate, irregular shell-like appearance which surrounds weak infrared
emission as seen in the 60$\mu$m IRAS data.  At a distance of 13 kpc
the size of the object is about 110$\times$220 pc and placed 500 pc
above the Galactic plane with a mass of 10$^5M_\odot$.  An expanding
shell with such a high mass and diameter cannot be explained by a
single supernova explosion or by a single stellar wind bubble.
We interpret the structure as a relic of
a distant stellar activity region powered by the joint action of
strong stellar winds from early type stars and supernova explosions.
\end{abstract}


\section{Introduction}

There is no generally accepted overall picture of the spiral arm
structure in the Galaxy, although we do know that there are star
forming regions beyond the Perseus arm \citep{woo71, sab86}. 
The \ion{H}{2} region BG 2107+49, for instance, is known to
be at R$_{\rm gal}$=14 kpc \citep{hig87} and \citet{chr79} found OB
stars at Galactocentric radii of R$_{\rm gal}$=15 to 25 kpc.
Furthermore, \citet{chi84}  and \citet{bli82} reported
\ion{H}{2} regions at R$_{\rm gal}\sim$18 kpc and \citet{dig94}
found molecular clouds associated with H$\bf\alpha$ emission at
R$_{\rm gal}$=18 to 28 kpc, whose kinetic temperature was comparable
to that of local star forming regions. One of the clouds they found,
designated as Cloud 2, has a kinematic distance of R$_{\rm gal}\simeq$28 kpc
\citep{geus93, dig94}.  Recently, \citet{kob00} have observed young
stellar objects associated with Cloud 2 and suggested a smaller
Galactocentric distance ($\sim20$ kpc). In addition recent data show that Cloud 2 is
associated with a larger \ion{H}{1} ring \citep[GSH~$138-01-94$]{sti01}.
These authors interpret this object as snowplowed material resulting
from a single supernova (SN) event.

The optical disk of the Galaxy may extend as far as 20 kpc from the
Galactic center.  The atomic gas, on the other hand, extends to a much
greater radius; R$_{\rm gal}\sim$30 kpc. Detection of star forming and
\ion{H}{2} regions beyond the optical disk provides information about
the composition of the Galactic disk.  The presence of these \ion{H}{2}
regions could explain the ionization of the diffuse gas at these large
distances and hence give the extent of the magneto-ionic medium in the
Galaxy. At these distances the only possible objects which may be
linked to the observed ionized gas are stars; because the general
interstellar radiation field is weaker, metallicity is lower
\citep{sha83, fic91} and there is less cosmic ray flux \citep{blo84}.

\ion{H}{2} regions far from the Galactic center are in fact a common
phenomenon in spiral galaxies similar to the Milky Way.  \citet{fer98}
have detected H$\alpha$ emission from star forming regions in the
extreme outer regions of the three nearby late-type spiral galaxies,
NGC~628, NGC~1058 and NGC~6946.  \citet{lel00} identified 137 {\sc
Hii} regions beyond a radius of $\sim16$ kpc with
narrow-band H$\alpha$ imaging in the galaxy NGC~628. M31 also has stellar
activity in its outer disk; \citet{cui01} found a population of
B stars correlated with the extended \ion{H}{1} distribution of M31.

With the increasing distance from the Sun, the spatial resolution of
the telescopes decreases. While single-antenna data can be used to
probe the \ion{H}{1} cloud concentrations \citep{dig94}, to resolve
these structures high resolution data are required. The new data at
$\sim1$ arcminute resolution provided by the Canadian Galactic Plane
Survey \citep[CGPS]{tay02} are well suited to study the structure and
dynamics of \ion{H}{1} regions at the edge of and even beyond the optical
disk.  Such a structure detected serendipitously south of the SNR
HB~21 in the CGPS database is the topic of this paper.  This paper
makes an attempt to clarify the possible origin of such large \ion{H}{1}
structures far from the Galactic center.  We analyze the \ion{H}{1}
spectral line data together with the available continuum data at 1420
MHz and infrared data to show that the detected structure is the result of
joint action of strong winds from early type stars and supernova (SN)
explosions.  We demonstrate that \ion{H}{1} spectral line observations
provide a very efficient method of tracing stellar activity at large
distances from the Galactic center, but not many such structures have
been detected because of the lack of data of sufficiently high
resolution covering large areas.
 
\section{\ion{H}{1} observations}

\ion{H}{1} line observations were carried out with the Dominion Radio
Astrophysical Observatory (DRAO) Synthesis Telescope \citep{lan00} as
part of the CGPS.  A detailed description of the data processing
routines can be found in \cite{wil99}.  The low spatial frequency {\sc
Hi} data are from the Low Resolution DRAO Survey of the CGPS region
observed with the DRAO 26-m Telescope \citep{hig00}. Parameters
relevant for the \ion{H}{1} data are given in Table~\ref{spec}.

\section{The \ion{H}{1} ring }

We have detected an \ion{H}{1} shell at $(\ell, b)=(90\degr, 2\fdg8)$ in
the \ion{H}{1} survey data of the CGPS (see Fig.~\ref{ch1} and
\ref{ch2}).  A shell-like structure appears in the foreground Perseus
arm gas at about $v_{LSR}=-$90 km s$^{-1}$. It slowly expands and gets
more prominent with higher negative velocities.  It is best pronounced
at a velocity of $v_{LSR}=-$99 km s$^{-1}$. At further negative
velocities it becomes smaller and fainter until it disappears at about
$v_{LSR}=-$114 km s$^{-1}$. We call this structure GSH~$90+03-99$,
in accordance with the current nomenclature for Galactic atomic
shells.

The atomic hydrogen map integrated between $v_{LSR}=-$94 and $-$106 km
s$^{-1}$ is given in Fig.~\ref{himap}.  The structure is about
$0\fdg$5$\times$1$\degr$ in size. The radial velocity of $-$99 km
s$^{-1}$ corresponds to a kinematic distance of 13 kpc, which would
give our \ion{H}{1} structure a linear size of 110$\times$220 pc, in
$\ell$ and $b$, according to a flat Galactic rotation curve with
R$_\odot=8.5$ kpc and $v_\odot=220$ km s$^{-1}$.  Toward the structure
there is very faint 1420 MHz continuum and infrared emission. In
Fig.~\ref{lb} we display the longitude-velocity diagram of the
structure. Behind the Perseus arm at the exact center of the detected
\ion{H}{1} structure we see a shell like feature moving toward us, while
the receding part is buried in the Perseus arm gas at lower negative
velocities.  This structure extends up to $v_{LSR}=-$114 km s$^{-1}$.
As a result the deduced expansion velocity of the \ion{H}{1} ring is
$v_{exp}\simeq$15 km s$^{-1}$.  The mass of the structure is about
10$^5$ M$_\odot$, which together with the expansion velocity gives a
kinetic energy of $2.3\times10^{50}$ erg currently in the expanding
shell, comparable to the explosion energy of an SNR. However, at such
a low expansion velocity the SNR should already be at the end of the
radiative phase with most of its energy  radiated away and
therefore no longer ``visible'' in continuum emission.

The kinematic age of the observed shell $t=\alpha ~R/v_{\rm exp}$,
with a mean radius $R=80$ pc,
would result in an age $1\times10^6$ yr for a radiative supernova
remnant ($\alpha=1/4$) and $3\times10^6$ yr for a supershell
($\alpha=3/5$). For GSH~$138-01-94$, these values are
$4\times10^6$ yr and $9\times10^6$ yr, respectively. Apparently the
object is too old to be an SNR, which  typically has an observable 
lifespan of  $10^5$ yr. This obtained age, on the other hand
is young in comparison to the distant Galactic shells listed by
\cite{hei84}, which are $\sim10^7$ yr old,  have  typical energies
$(1-5)\times10^{52}$ erg and are usually up to 10 times larger
than the detected structure.  Nevertheless, the object has
a mass comparable to that of distant Galactic shells and yet
a size comparable to that  of local  shells.
For instance, the Local Bubble, North Polar Spur and Gum Nebula
are just twice as large as this \ion{H}{1} shell \citep{hei98}.

We have checked the region in the Columbia CO survey \citep{dam87}.
There is no evidence of associated molecular material.  However this
survey has coarse angular resolution and is undersampled.  Especially
the area of interest has a low signal-to-noise ratio; a signal below
0.5 K could not be detected.  The maximum molecular hydrogen column
density would be $N_{H_2} \simeq 1.2\times10^{20}$ cm$^{-2}$ per km
s$^{-1}$.  Assuming a molecular material distribution comparable to
the atomic hydrogen distribution, a maximum molecular mass of
$M\simeq10^5 ~M_\odot$ in the velocity interval, shown in Fig.~1, for
instance, could stay undetected.  Thus, non-detection in the Columbia
data does not necessarily mean that there is no molecular gas in the
region.

\subsection{Would it be possible to identify any related stars?}

\ion{H}{2} regions are expected to include stars. However, stellar
light suffers from absorption and reddening by the intervening
material.  Therefore, beyond a certain distance stars of an \ion{H}{2}
region cannot be observed.

At the north-western part of the structure there is the UV star
Lan~110 \citep{lan94}, at $\ell=89\fdg9$ and $b=2\fdg91$.  We have
calculated the \ion{H}{1} absorption profile of a nearby extragalactic
source at $(\ell, b)= (89\fdg93, 3\fdg05)$ and thereby found a
foreground \ion{H}{1} column density of 9.6$\times$10$^{21}$ cm$^{-2}$.
Combining that with the observed optical parameters of the star
results in a color of $(U-B)_0 =-$1.85.  An O3 star would have
$(U-B)_0=-$1.22.  Therefore, this UV star must be located in front of
the Perseus arm rather than behind it. This star is apparently not
related to the detected structure. This also demonstrates that the
detection of stars within the structure would be difficult.

To clarify whether massive stars in the \ion{H}{1} structure are
observable, we have calculated extinction and reddening of typical O
and early B type stars, based on the foreground \ion{H}{1} column
density and the distance.  The measured foreground \ion{H}{1} 
column density is 9.6$\times10^{21}$ cm$^{-2}$, which  corresponds
to a reddening of
\begin{equation}
E_{\rm B-V}= \frac{ N_{HI}  }
      { 4.8\times10^{21} \mbox{cm}^{-2} } = 2.0 ~\mbox{mag}.
\end{equation}
as given by \citet{boh78}.  This gives a visual extinction of $A_V =
6.4$ magnitude.

A typical, main-sequence O3 star has an absolute magnitude $M_V =
-6.0$ and color indices $(B-V)_{0} =-0.33$ and $(U-B)_{0} =-1.22$. The
above reddening and extinction in turn correspond to  apparent
visual magnitudes of $m_V=16^{\rm m}$, $m_B=17.7^{\rm m}$ and
$m_U=18^{\rm m}$.

For a main-sequence B1 star, with $M_V = -3.2$ and color indices
$(B-V)_{0} =-0.27$ and $(U-B)_{0} =-0.95$, the above calculation
yields visual magnitudes $m_V=18.8^{\rm m}$, $m_B=20.5^{\rm m}$ and
$m_U=21.2^{\rm m}$.  We note that, for both O3 and B1 stars the observed
color indices are $(B-V)>0$ and $(U-B)>0$.  \citet{lan94} have
detected stars in the range from $(U-B) =0 $ to $-1.5$, and in
magnitude from $m_U=10^{\rm m}$ to $21^{\rm m}$ in this
region. Therefore, stars with positive $(U-B)$, as those which were
mentioned above, are missing in their list.

These calculations show that deep measurements have to be made in
order to detect early type stars which may contribute to the
energy budget of the \ion{H}{1} ring by their stellar winds.

Near infrared measurements, on the other hand, might provide
further hints about the existence of embedded, young OB clusters
inside the \ion{H}{1} structure like those found in
GSH~$138-01-94$ \citep{kob00}, since there exists an
infrared source in the area enclosed by the \ion{H}{1} shell
(see Sect.~\ref{sec4}).
However, such sensitive near infrared
data toward GSH~$90+03-99$ are currently not available. 

\section{Other objects in the region} \label{sec4} 

The infrared source IRAS 20565+5003 is located almost at the center of
the \ion{H}{1} structure.  According to \citet{bro96} the flux ratios
25$\mu$m/12$\mu$m ($>3.7$) and 60$\mu$m/12$\mu$m ($>19.3$) indicate a
compact \ion{H}{2} region. The observed infrared intensities for our
source (0.91, 3.50 and 31.57 Jy, at 12, 25 and 60 $\mu$m,
respectively) make this source a strong candidate for an \ion{H}{2}
region.  \citet{rud96}  observed 10 IRAS point sources, at
distances R$_{\rm gal}$ 15 to 18 kpc, with the VLA at 2 and 6 cm. They
found the spectral indices of these sources are consistent with the
optically thin free-free emission from \ion{H}{2} regions, signaling
stellar activity at large distances.  The distance of IRAS 20565+5003
is unknown and we have only circumstantial evidence that it is located
within the \ion{H}{1} ring. Therefore an association between the two is
not certain, although the positional coincidence is attractive and
compelling.  Other IRAS sources in the region are away from the {\sc
Hi} structure or do not satisfy the required infrared ratio criteria.

The \ion{H}{2} region BFS~4 at $(\ell, b)=90\fdg32, +2\fdg67$ is also
in the region. However, the observed radial velocity, 1.1$\pm$0.5 km
s$^{-1}$ \citep{bli82}, of BFS~4 implies that it is a local object.

\section{SNR or  OB association?}

The observed \ion{H}{1} structure may be attributed to an old SNR, which
has already cooled to temperatures where ionized hydrogen recombines
and creates the observed \ion{H}{1} shell.  Assuming the structure is
the result of such an event we can calculate the maximum radius of the
SNR as a function of the explosion energy and the ambient density. The
mass of the observed \ion{H}{1} structure is about 10$^{5}$~M$_\odot$,
comparable to a typical giant molecular cloud, assuming the gas is
optically thin.  Thus, we obtain n$_0\simeq$2.0 cm$^{-3}$ before the
structure was formed.  Due to the high Galactocentric radius of the
\ion{H}{1} ring, we have to take the metallicity into account, because
lower metallicity causes slower cooling, which in turn results in a
lower energy loss rate. Hence in the outer Galaxy SNRs live
longer. There are two recent publications dealing with the dynamics of
SNRs at late stages of their evolution taking metallicity effects into
account \citep{cio88, tho98}, latter based on the central equations of
the former.  Both discuss the so-called radiative expansion or
snowplow phase at the end of SNR evolution. This phase consists of two
parts: the cooling dominated first part and later the merger with the
interstellar medium. At the end of the first part the SNR is no longer
distinguishable from the environment. Thus we adopt the radius at the
end of the first part as the maximum observable radius of an SNR.
Following \citet{cio88} the maximum radius is
\begin{equation}
R_{\rm merge} = 51.3 ~E_0^{31/98} ~ n_0^{-18/49} \zeta^{-5/98}
\label{merge}
\end{equation}
where $E_0$ is the explosion energy in units of 10$^{51}$ erg, $n_0$
is the mean ambient particle density in cm$^{-3}$, and $\zeta$ is the
metallicity normalized to the solar value.

Using this model we can estimate the energy required if the ring is
the result of a single SN explosion. Since Eqn.~\ref{merge} provides
the maximum radius, the corresponding explosion energy will be a lower
limit for the energy requirement. Results obtained from this
calculation for GSH~$90+03-99$ and GSH~$138-01-94$, discovered by
\citet{sti01}, are given in Table~\ref{pds}.

In order to reduce the required explosion energy the SNR would have to
be closer.  We have calculated the mass of the structure from the
observed hydrogen column density.  Therefore it is proportional to
the square of the distance, implying a linear dependence of $n_0$ on
distance, $d$.  Combining this with Eqn~\ref{merge} we obtain
$E_0 \propto d^2$.  The maximum distance, $d_{\rm max}$, we would
obtain for an explosion energy of $10^{51}$ erg is also given in
Table~\ref{pds}.

Consequently, for the single SNR explanation the maximum distance of
the structures, as calculated above put these objects significantly
closer.  The systemic velocities corresponding to the $d_{\rm max}$
values for these two objects lie between $-20$ and $-40$ km s$^{-1}$
for GSH~$90+03-99$ and $-40$ and $-55$ km s$^{-1}$ for
GSH~$138-01-94$.  These velocities, however, contradict the observed
systemic values and require extremely high random motions never
observed before. Therefore, we find such a close distance, hence a
single SNR origin, for both of the objects {\it highly} unlikely. We
should emphasize that all calculated values are extrema, making the
single SNR hypothesis even more unlikely.

A single O type or an early B type star would create a stellar wind
bubble of radius \citep{mckee84} 
\begin{equation}
r_w = 56 ~n_0^{-0.3},
\end{equation}
which in turn gives about 45 pc for the \ion{H}{1} ring, with the above
ambient density.  The corresponding radius of a wind bubble for
GSH~$138-01-94$, is 83 pc.  Therefore an association of several
massive stars would be sufficient to form such bubbles.  By comparing
the SNR and wind-bubble calculations above, we conclude that a
structure formed by a SN explosion and a stellar wind bubble would
have comparable characteristics. Thus, the shells discussed in this
paper could be formed by approximately equal numbers of SN explosions
or winds of massive stars.

We find that a single SN event or a stellar wind bubble of a single
massive star cannot be held responsible for the formation of such big
structures. An OB association hidden within the structure, however,
would create enough stellar wind power to carve such a region.
Additionally, one must take into account the fact that these young
stars will sometime become SN, contributing to the energy budget of
the big \ion{H}{1} structure. Stars are usually formed in groups and the
creation of a single and yet massive star alone seems to be
improbable.
 
We note that both of the \ion{H}{1} structures discussed here have
comparable characteristics.  One exception is that GSH~$138-01-94$ was
first detected as an \ion{H}{1} concentration from the Maryland-Green
Bank Survey \citep{wes82}, later detected in CO
emission \citep{dig94}. An early type B-star is reported to be
associated with it \citep[``MR~1'']{muz74}. The radial velocity
of  MR~1 \citep[$-90\pm13$ km s$^{-1}$]{sma96} is comparable
to the velocity of GSH~$138-01-94$.  Finally 
\citet{geus93} and \citet{kob00} confirmed the existence of the ionizing
star and classified the object as an \ion{H}{2} region.  Such
associated objects are not known towards GSH~$90+03-99$, since
there are no such deep measurements.  However its similarity to
GSH~$138-01-94$ strengthen the hypothesis that it contains stellar
activity.

\section{Summary}
We reported the detection of an expanding hydrogen shell
(GSH~$90+03-99$) most likely powered by a group of massive stars in
its center. We interpret GSH~$138-01-94$ recently discovered by
\citet{sti01} as another example of a SN/OB-association powered 
bubble at a large distance.
 We think this is a common phenomenon and we expect the
detection of similar objects in the future.

 The size of GSH~$90+03-99$ is 110$\times$220 pc at a distance of 13
kpc. It contains 10$^5$ M$_\odot$ of neutral material and is expanding at 15
km s$^{-1}$. This gives a kinetic energy of $2.3\times10^{50}$ erg.
If OB associations are sufficiently distant, it becomes difficult to
detect the optical emission from their members.  In this case
atomic and molecular gas are their best tracers and a survey with high
angular resolution, covering a large area of the Galactic plane is
ideal for their detection.  Spectral information is decisive to
distinguish the SNRs from \ion{H}{2} regions; but at large distances,
due to low surface brightness, this information is usually
unavailable.  Therefore, energetics of the observed structures become
important to characterize these objects.  \ion{H}{1}, being unaffected by
interstellar absorption, is an easily detectable tracer of objects at
large distances, and is also very efficient for the study of their
kinematics.

\acknowledgments{
 We thank Tom Landecker and Charles Kerton for
careful reading of the manuscript and discussions.
The Dominion Radio Astrophysical Observatory is a
National Facility operated by the National Research Council.  The
Canadian Galactic Plane Survey is a Canadian project with
international partners, and is supported by the Natural Sciences and
Engineering Research Council (NSERC).
This research made use of SIMBAD database.
}

\clearpage
\begin{table}
\caption{Specifications of \ion{H}{1} spectral line data \label{spec}}
\begin{tabular}{ll}
\hline
Parameter                       &  Value                        \\ \hline
Polarization                    &  Left and right circular      \\
Bandwidth                       &  1  MHz                       \\
Field size (to 20\% response)   &  2.6   degree                 \\
Synthesized beam (EW$\times$NS) &  1$\times$ 1 arcmin           \\
Velocity resolution             &  0.824 km s$^{-1}$            \\
Total 256 Channel span          &  $-$158 to $+$52 km s$^{-1}$  \\ \hline
\end{tabular}
\end{table}

\begin{table}
\caption{Characteristics of the two \ion{H}{1} rings calculated from
Eqn.~\ref{merge}. Here $\zeta$ is the ratio of metallicity to the
solar value, $R_{\rm merge}$ is the maximum radius of a SNR before it
disappears assuming the explosion energy is $10^{51}$ erg, $E_0$ is
the explosion energy required when the observed radius of the object
is equal to $R_{\rm merge}$, and $d_{\rm max}$ is the distance to the
object if it is due to a single SNR with an explosion energy of
$10^{51}$ erg.
\label{pds}
}
\begin{tabular}{lrrrr} \hline
\multicolumn {1}{c}{Object}          &
 \multicolumn {1}{c}{$\zeta$}         &
  \multicolumn {1}{c}{$R_{\rm merge}$} &
   \multicolumn {1}{c}{$E_0 $ }         &
    \multicolumn {1}{c}{$d_{\rm max}$}   \\
\multicolumn {1}{c}{ }               &
 \multicolumn {1}{c}{ }               &
  \multicolumn {1}{c}{(pc)}            &
   \multicolumn {1}{c}{($10^{51}$ erg)} &
   \multicolumn {1}{c}{(kpc)}           \\ \hline

GSH~$138-01-94$  & 0.1     & 93.3  &   8.0  & 5.9 \\
                 & 1.0     & 83.0  &  11.6  & 4.9 \\
GSH~$ 90+03-99$  & 0.1     & 44.7  &   6.3  & 5.2 \\
                 & 1.0     & 39.8  &   9.1  & 4.3 \\
\hline
\end{tabular}
\end{table}

\begin{figure*}\centering
\caption[fig1]{\ion{H}{1} channel maps toward the \ion{H}{1} shell.
The contour represents the 1420 MHz radio continuum emission
at  7.8 K T$_{\rm B}$.
\label{ch1} }
\end{figure*}

\begin{figure*}\centering
\caption[fig2]{Same as Fig.~\ref{ch1} but for different
radial velocities.
\label{ch2} }
\end{figure*}

\begin{figure*}\centering
\caption[fig3]{
\ion{H}{1} emission  toward  the detected  ring  of  neutral
hydrogen integrated between $-$94 and $-$106 km s$^{-1}$
\label{himap} }
\end{figure*}

\begin{figure*}\centering
\caption[fig4.eps]{
Longitude--velocity diagram  of the detected \ion{H}{1} structure.
The arrow marks the approaching shell.
\label{lb}}
\end{figure*}

\label{lastpage}
\end{document}